\documentclass[aps,prb,superscriptaddress, sd,amsmath,reprint,longbibliography,amssymb]{revtex4-1}

\usepackage[english]{babel}

\usepackage[letterpaper,top=2cm,bottom=2cm,left=3cm,right=3cm,marginparwidth=1.75cm]{geometry}

\usepackage{amsmath}
\usepackage{graphicx}
\usepackage{xcolor}
\usepackage[colorlinks=true, allcolors=blue]{hyperref}
\usepackage[utf8]{inputenc}
\begin{document}

\title{Size effects in the Verwey transition of nanometer-thick micron-wide magnetite crystals}

\author{Adolfo del Campo}
\affiliation{Instituto de Cerámica y Vidrio, CSIC, Madrid E-28049, Spain}
\author{Sandra Ruiz-G\'omez}
\affiliation{Max-Planck-Institut für Chemische Physik fester Stoffe, 01187 Dresden, Germany}
\author{Eva M. Trapero}
\affiliation{Instituto de Qu\'{\i}mica F\'{\i}sica ``Rocasolano'', CSIC, Madrid E-28006, Spain}
\author{Cecilia Granados-Miralles}
\affiliation{Instituto de Cerámica y Vidrio, CSIC, Madrid E-28049, Spain}
\author{Adrián Quesada}
\affiliation{Instituto de Cerámica y Vidrio, CSIC, Madrid E-28049, Spain}
\author{Michael Foerster}
\affiliation{Alba Synchrotron Light Facility, 08290, Cerdanyola del Valles E-08290, Spain}
\author{Lucía Aballe}
\affiliation{Alba Synchrotron Light Facility, 08290, Cerdanyola del Valles E-08290, Spain}
\author{José Emilio Prieto}
\affiliation{Instituto de Qu\'{\i}mica F\'{\i}sica ``Rocasolano'', CSIC, Madrid E-28006, Spain}
\author{Juan de la Figuera}
\affiliation{Instituto de Qu\'{\i}mica F\'{\i}sica ``Rocasolano'', CSIC, Madrid E-28006, Spain}
\email{juan.delafiguera@iqfr.csic.es}
\date{\today}



\begin{abstract}
We have monitored the Verwey transition in micrometer-wide, nanometer-thick magnetite islands on epitaxial Ru films on Al$_2$O$_3$(0001) using Raman spectroscopy. The islands have been grown by high-temperature oxygen-assisted molecular beam epitaxy. Below 100~K and for thicknesses above 20~nm the Raman spectra correspond to those observed in bulk crystals and high quality thin films for the sub-Verwey magnetite structure. At room temperature the width of the cubic phase modes is similar to the best reported in bulk crystals,
indicating a similar level of electron-phonon interaction. The evolution of the Raman spectra upon cooling suggests that for islands thicker than 20~nm, structural changes appear first at temperatures starting at 150~K and the Verwey transition itself takes place at around 115~K. However, islands thinner than 20 nm show a very different Raman spectra indicating that while a transition takes place, the charge order of the ultrathin islands differs markedly from their thicker counterparts.
\end{abstract}

\maketitle

Magnetite is a mixed valence iron oxide with a cubic, inverse spinel crystalline structure at room temperature. Magnetite undergoes a phase transition, the Verwey transition\cite{Verwey}, upon cooling at a temperature of 120~K for bulk crystals. Below the phase transition the resistivity increases by two orders of magnitude, the magnetic anisotropy increases and the crystallographic structure changes to monoclinic. The origin of these changes has been debated for a century and this has pushed forward developments in solid state physics such as the study of the Mott transition. However, many details of the transformation, such as the symmetry of the unit cell or the type of charge order at low temperatures have been the subject of heated debates\cite{walz_verwey_2002,garcia_verwey_2004}. Lately there is consensus on a detailed atomic model of the low temperature phase based on high resolution x-ray diffraction data\cite{SennNature2012}. 
From this model, a complex charge order has been suggested, composed of an arrangement of so-called trimeron units: Fe$^{3+}$-Fe$^{2+}$-Fe$^{3+}$ linear chains. However, many details of the phase transition continue to be debated. In particular, given the trend towards the study of materials in nanostructure form in general and for applications in spintronics of magnetite in particular, the effect of particle size on the Verwey transition is still an open question. Several studies indicate that for nanoparticles smaller than 20 nanometers the transition is suppressed and disappears completely for sizes smaller than 6~nm\cite{LeeNanoL2015}. Size effects such as a decrease of the Verwey temperature have also been detected in films for thicknesses below 60~nm\cite{KoreckiThinSolF2002}.

The Verwey phase transition can be followed using very different techniques. However, most of them are not sensitive to a particular charge order but are rather more indirect. For example, the measurements of resistivity or magnetization changes are examples of averaging techniques, where the information arises from large ensembles of atoms. Other techniques are of local character, such as Mössbauer spectroscopy or Raman spectroscopy. However, Mössbauer spectroscopy, due to the low signal to noise ratio, is not well suited to study nanostructures. In contrast, Raman spectroscopy has the advantage that it can be applied to small areas of a sample. The usefulness of Raman spectroscopy to follow the Verwey transition was already shown in the classic work of J.L Verble\cite{VerblePRB1974}. Raman spectroscopy continues to be applied to study the Verwey transition on the surface of bulk crystals\cite{GasparovPRB2000,GuptaPRB2002,GasparovPRB2007}. Often sharp changes in some Raman modes is detected, and an increase of the background intensity\cite{GasparovPRB2007}, which was attributed to the opening of the band gap observed in photoemission spectroscopy\cite{ParkPRB1997}. More recently, Raman spectroscopy has been used to study magnetite films grown on several oxide \cite{PhaseJAP2006,YazdiNJP2013,KumarPRB2014} and metal\cite{LewandowskiSCT2015} crystals. 

In the present work, we report on the observation with Raman spectro-microscopy of the Verwey transition of magnetite micro-crystals grown on thin epitaxial Ru films. The magnetite crystals possess a high quality as judged from their structural and magnetic properties: they have mostly triangular shapes with typical lateral sizes of several micrometers and heights ranging from a few nanometers up to a hundred nanometers and they present an excellent order as evidenced by low-energy electron diffraction. Their magnetic properties resemble those of bulk magnetite\cite{SandraNanoscale2018} and present well defined magnetic domains already for nanometer thicknesses\cite{MontiPRB2012}.

\section{Methods}

The Ru films were deposited on Al$_2$O$_3$(0001) single crystal substrates in a home-built magnetron sputtering system. A typical DC magnetron power used was 20~W after 10~min pre-sputtering of a 2" Ru target from Evochem GmbH. The substrates were kept at 900~K during film growth. A detailed characterization of the films by Rutherford Backscattering Spectroscopy and channeling has been recently described\cite{PrietoASS2022}. The films were then transferred to an ultrahigh vacuum chamber containing a low-energy electron microscope and annealed at temperatures of up to 1300~K. Both the Elmitec III LEEM at the IQFR and the Elmitec SPELEEM at the CIRCE beamline of the ALBA Synchrotron\cite{CIRCE} were used for the growth of the magnetite islands.  The growth was performed by introducing a pressure of 10$^{-6}$ mbar of molecular oxygen while the substrate was kept at a temperature of 800$^\circ$C and Fe was deposited from a home-made doser consisting of a Fe rod 5~mm in diameter in a water jacket heated by electron bombardment from a W filament with a typical power of 25~W. X-ray based characterization using photoemission microscopy was performed at the  ALBA Synchrotron CIRCE beamline.

The Raman spectra were acquired with a commercial Renishaw Witec Alpha 300RA confocal Raman spectrometer, using a 20$\times$ objective with a numerical aperture of 0.4. The light source was a 532~nm laser operated at 1~mW power, selected in order to avoid oxidation of the samples. The spectra presented are the average of 5 scans, each acquired with 30~s integration time. Measurements were performed at different temperatures between room temperature and liquid nitrogen temperature (77~K). The focus was adjusted at each temperature. We have fitted the Raman spectra by a sum of Lorentzians including a third degree polynomial for the background after applying a fifth order median filter. 

\section{Results and discussion}

\begin{figure}[htb]
	\centerline{\includegraphics[width=0.45\textwidth]{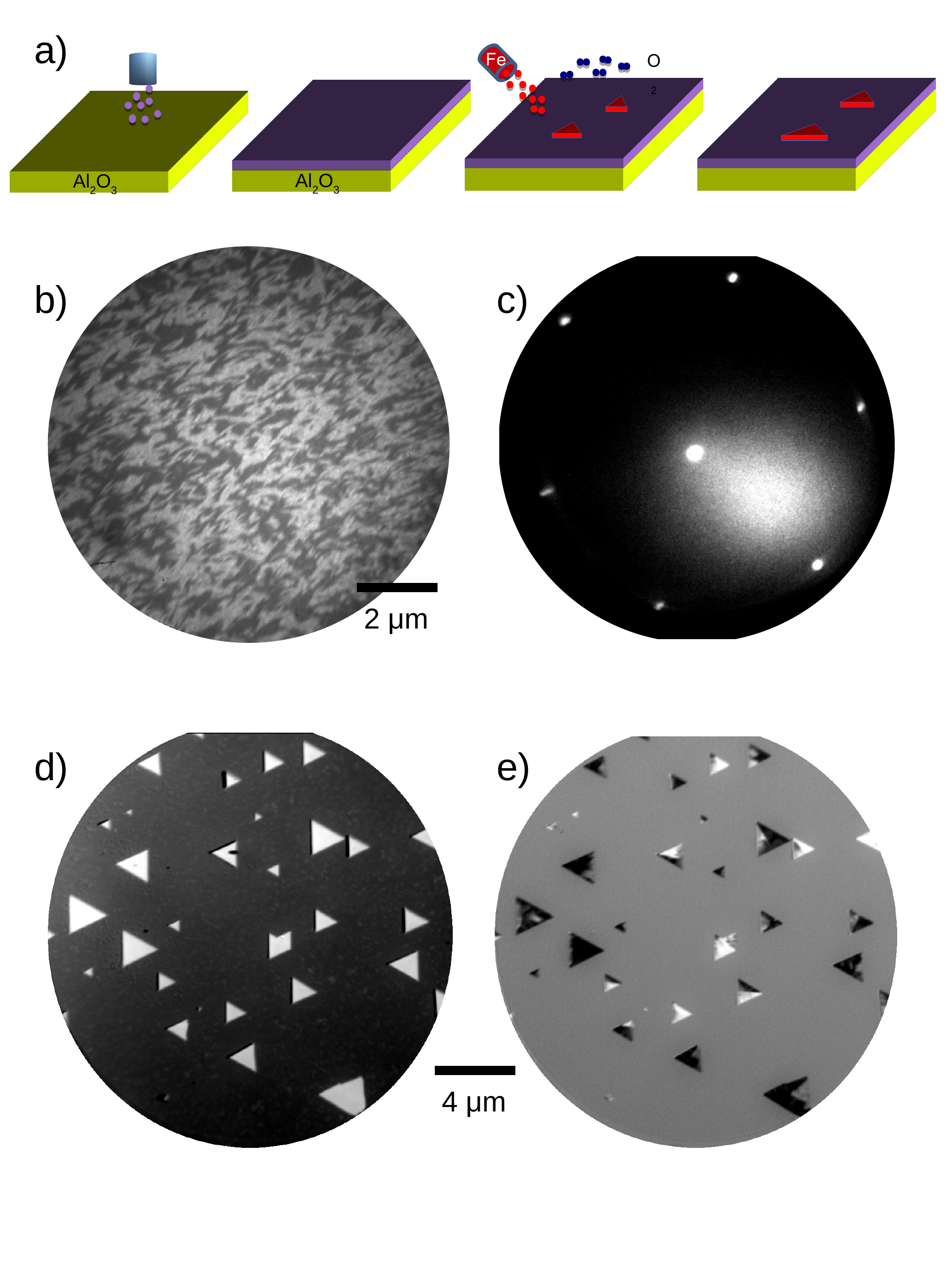}}
	\caption{Growth of the magnetite micro-crystals. a) Schematics of the procedure, showing the growth by sputtering of the Ru film and the subsequent growth of the magnetite islands on top by high-temperature oxygen-assisted molecular beam epitaxy. b) Low-energy electron microscopy image showing the Ru film surface in dark field mode (successive atomic terrace show alternately dark - light gray contrast). c) Low-energy electron diffraction pattern of the Ru film. d) X-ray absorption image acquired at the white line of the Fe L$_3$ absorption edge. e) x-ray magnetic circular dichroism image of the same region.}
	\label{fg:growth}
\end{figure}

The sample growth process is schematically depicted in Figure~\ref{fg:growth}a; it is similar to the procedure employed by Flege and coworkers to grow ceria islands on Ru films\cite{FlegeCGD2016}. The Ru films were grown by magnetron sputtering on Al$_2$O$_3$(0001) single crystals. Despite having a substantial density of steps compared with the Ru single crystals employed in Ref.~\onlinecite{SandraNanoscale2018}, as shown in Figure~\ref{fg:growth}b, the films are well ordered and single crystalline as previously characterized\cite{PrietoASS2022} and as shown by the diffraction pattern in Figure~\ref{fg:growth}c. The magnetite crystals themselves were grown on top of the Ru films by oxygen-assisted high-temperature molecular beam epitaxy under {\it in-situ} observation by low-energy electron microscopy. 
The growth of iron oxide proceeds in the same way as for single crystal Ru(0001) substrates\cite{SantosJPC2009}; for this case, it has been characterized  by microspot low-energy electron diffraction, as well as x-ray photoemission, x-ray absorption spectroscopy and x-ray magnetic circular dichroism in photoemission microscopy\cite{MontiPRB2012,MontiJPC2012,SandraNanoscale2018}. The crystals grow on top of a wetting layer composed of two atomic layers of FeO\cite{IreneJPc2013} that first covers the Ru substrate and then 3-dimensional islands of magnetite nucleate and grow on top. An x-ray absorption image acquired at the Fe L$_3$ edge is shown in Figure~\ref{fg:growth}d. The crystals appear as white triangles, as they contain more Fe than the surrounding FeO film. X-ray magnetic circular dichroism images reveal the magnetic domains of the islands (Figure~\ref{fg:growth}e), while the FeO areas in-between are not magnetic at room temperature.

\begin{figure}[htb]
	\centerline{\includegraphics[width=0.45\textwidth]{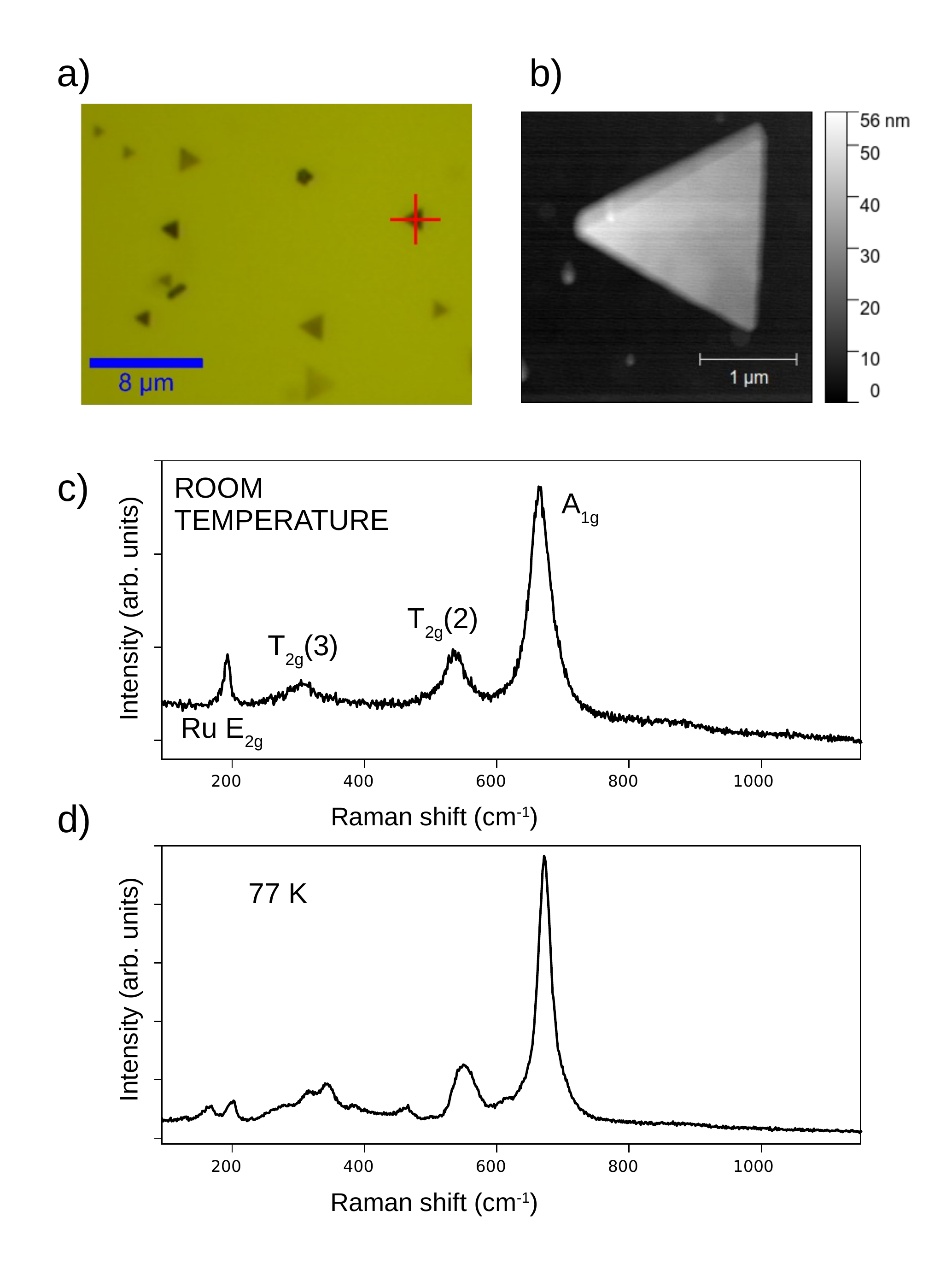}}
	\caption{a) Optical microscopy image of magnetite triangles on a Ru film. b) Atomic force microscopy image of one of the islands. c) and d) Raman spectra acquired on the island shown in b) at room and low temperature, respectively.}
	\label{fg:overview}
\end{figure}

The film was then taken out of ultra-high-vacuum and studied {\it ex-situ} by confocal microscopy and Raman spectroscopy. In Figure~\ref{fg:overview}a we show an optical micrograph of a region containing a few magnetite crystals. Atomic force microscopy of one of the islands, marked with a red cross, reveals that across its triangular shape, with 2~$\mu$m side, the height varies from 30 to 50~nm. The room temperature and low temperature (77~K) Raman spectra acquired on the island are shown in Figure~\ref{fg:overview}c,d respectively. 

Magnetite has at room temperature the cubic inverse spinel structure, which corresponds to the $Fd3m$ space group. All the Fe$^{2+}$ cations and half of the Fe$^{3+}$ cations are located at octahedral sites, while the remaining Fe$^{3+}$ cations occupy tetrahedral sites. The rombohedral unit cell contains only 14 atoms: 2 tetrahedral iron, 4 octahedral iron and 8 oxygen atoms. A group theoretical analysis of the structure predicts 42 vibrational modes, of which 5 are Raman active, those with symmetries $A_{1g}$, $E_g$ and the three different $T_{2g}$\cite{WhiteSA1967}. All these modes correspond to breathing modes of the tetrahedral cations. The most intense mode is the $A_{1g}$ one, observed at $\omega= 665~$cm$^{-1}$, which corresponds to the symmetric breathing of the oxygen anions around each tetrahedral cation. The other peaks found for magnetite are two of the $T_{2g}$ ones, at 310 and 535 cm$^{-1}$, respectively. In agreement with other studies of thin films\cite{YazdiNJP2013} we do not observe either of the other two possible modes, a $T_{2g}$ one and the $E_g$ one. The large peak at 192 cm$^{-1}$ can be assigned to the Ru film, as it is detected also in regions were no magnetite crystals are observed. It corresponds to the Ru $E_{2g}$ transverse optical phonon arising from the shear of the two sublattices of the hcp unit cell\cite{GrimsditchPRL1996}. Such a peak has been detected in Ru films and multilayers\cite{GrimsditchPRL1996,chung_thermally_2008} as well as in other hcp metals\cite{DemersPRB1988}. 

\begin{table}[]
    \centering
    \begin{tabular}{c | c c c}
         Raman mode & $A_{1g}$ & $T_{2g}(2)$ & $T_{2g}(3)$ \\
         \hline \hline
         $\omega$ (cm$^{-1}$) & 664.7$\pm$0.7 & 534$\pm$2 & 311$\pm$10 \\
         $\Gamma$ (cm$^{-1}$) & 38$\pm$1 & 46$\pm$6 & 81$\pm$45 \\
         $\Gamma/\omega^2$ (eV$^{-1}$) & 0.69$\pm$0.02 & 1.3$\pm$0.2 & 7$\pm$3 \\
         $\lambda$ & 0.037$\pm0.001 $ & 0.14$\pm$0.02 & 1.05$\pm$0.5 \\
         \hline
    \end{tabular}
    \caption{Wavenumbers $\omega$ (cm$^{-1}$), FWHM $\Gamma$ (cm$^{-1}$), $\Gamma/\omega^2$ (eV) and strength $\lambda$ of the electron-phonon interaction estimated from the Raman peaks of each mode measured at room temperature, averaged over several crystals of micrometric width and height in the range 10~nm - 60~nm.}
    \label{tab:phonon_interaction}
\end{table}

The spectra of other islands is very similar at room temperature. The FWHM ($\Gamma$) of the modes shown in Figure~\ref{fg:overview}c are respectively 38, 41, 118 and 11 cm$^{-1}$ for the magnetite $A_{1g}$, $T_{2g}(2)$, $T_{2g}(3)$ and the Ru $E_{2g}$ modes. The spinel modes are much wider than, for example, the one of the underlying Ru film. This has been noted in all previously published studies of magnetite. The main contribution to this larger width has been attributed by Verble\cite{VerblePRB1974} to electronic disorder from the random arrangement of Fe$^{2+}$ and Fe$^{3+}$ ions on the B sites. Gupta et al.\cite{GuptaPRB2002} explained it in terms of a strong electron-phonon interaction related to the decay of phonons into electron-hole pairs. In such case, the strength of the electron-phonon interaction at room temperature (i.e. in the disordered state with Fe$^{2+}$ and Fe$^{3+}$ ions on the B sites) can be estimated from the width of a given Raman mode according to:
\[
\frac{\Gamma}{\omega^2} = \frac{2\pi}{g}\lambda N(E_F)
\]
where $g$ is the degeneracy of the mode, and $\lambda$ is the intensity of the electron-phonon interaction. We have measured these values for islands with thickness between 10~nm and 60 nm and found no clear dependence on the island height. This suggests that the short range order does not differ appreciably between islands. We present in table~\ref{tab:phonon_interaction} the average values of the relevant parameters together with their statistical errors, determined from six different islands with heights in the 10--60 nm range.
In table~\ref{tab:comparison}, we include a comparison of our estimations for the electron-phonon interaction strength with published results.

The electron-phonon interaction is much weaker for the $A_{1g}$ mode than for the $T_{2g}(2)$ one, which in turn is an order of magnitude smaller than for the  $T_{2g}(3)$ mode. We note that our results are comparable to those reported for the best films\cite{YazdiNJP2013}, where it is mentioned that the samples were selected for their high Verwey temperature. That the interaction with the $T_{2g}$ modes is much higher has been suggested to be due to sharing the symmetry of electronic states at the Fermi level\cite{GuptaPRB2002}. However, there is no clear explanation for the reason why it should be so different between the two $T_{2g}$ modes. Phase el al.\cite{PhaseJAP2006} attributed it to the presence of antiphase boundaries. While we expect that our crystals, having grown each presumably from a single nucleus, lack antiphase boundaries, we note that the films reported by Yazdi\cite{YazdiNJP2013} should contain a large number of antiphase boundaries, but they present (together with our results) the lowest values for thin films.

\begin{table}[]
    \centering
    \begin{tabular}{c | c c c }
          & $\lambda_{A_{1g}}$ & $\lambda_{T_{2g}(2)}$ & $\lambda_{T_{3g}(3)}$ \\
         \hline \hline
          \multicolumn{4}{c}{This work} \\
         Ru/Al$_2$O$_3$ & 0.037$\pm0.001$ & 0.14$\pm$0.02 & 1.05$\pm$0.5 \\
         \hline
          \multicolumn{4}{c}{Bulk single crystal} \\
         \hline
         Ref.~\onlinecite{VerblePRB1974} & 0.034$^{*}$ & & \\
         Ref.~\onlinecite{GasparovPRB2000} & 0.038$^{*}$ & & \\
         Ref.~\onlinecite{GuptaPRB2002} & 0.045 & 0.20 & 0.51 \\
         \hline
          \multicolumn{4}{c}{Thin films} \\
         \hline
         MgO(100)\cite{PhaseJAP2006} & 0.047 &  0.33 & 0.98 \\
         MgO(100)\cite{YazdiNJP2013} & 0.035$^{*}$ & 0.11$^{*}$ & 1.06$^{*}$ \\
         Al$_2$O$_3$(0001)\cite{YazdiNJP2013} & 0.035$^{*}$ & 0.13$^{*}$ & 1.01$^{*}$ \\
         TiN/Si(100)\cite{KumarAPL2013} & 0.039$^{*}$ & 0.456 & \\
         \hline
    \end{tabular}
    \caption{Values for the electron-phonon interaction strength for the main three modes reported in the literature. The ones marked with an asterisk have been estimated from the data provided in the corresponding reference.}
    \label{tab:comparison}
\end{table}

The spectrum at low temperature (Figure~\ref{fg:overview}d) shows many more peaks. This is a distinctive feature of the Verwey transition in magnetite\cite{VerblePRB1974}. The spectra of the islands are very similar to those reported below the Verwey transition both for magnetite single crystals\cite{VerblePRB1974,GasparovPRB2000,GasparovPRB2007} and for high quality films\cite{PhaseJAP2006,YazdiNJP2013}. We take this as proof of the occurrence of the Verwey transition in our micrometric-wide, nanometer-thick magnetite islands.

\begin{figure}[htb]
	\centerline{\includegraphics[width=0.45\textwidth]{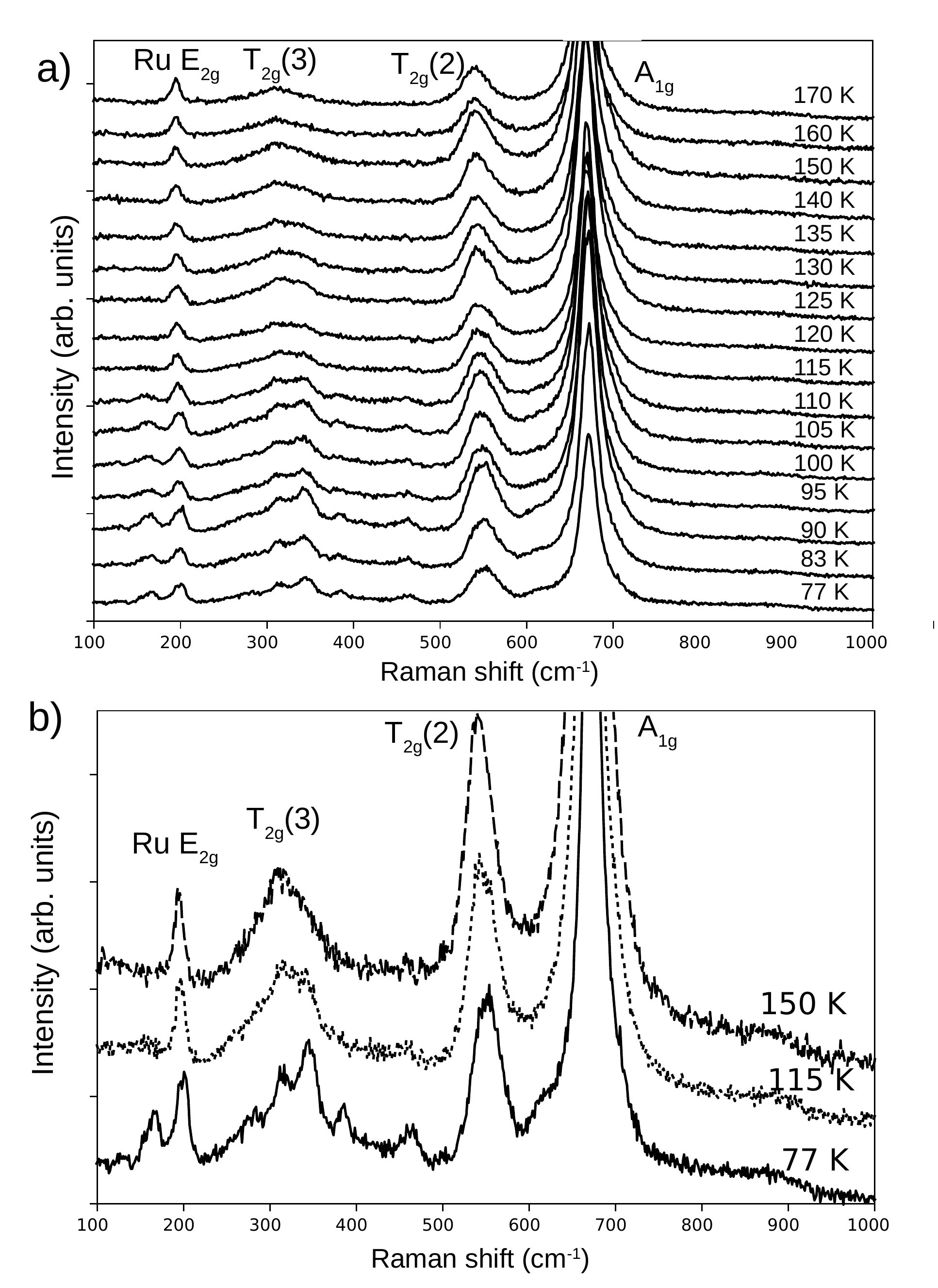}}
	\caption{a) Raman spectra acquired through the Verwey transition in the range 170~K - 77~K. b) Raman spectra at three representative temperatures to show the characteristics of the evolution with temperature (150~K, 115~K, and 77~K), acquired on the island displayed in the previous figure. }
	\label{fg:temp}
\end{figure}

The detailed temperature evolution of the Raman spectrum of the island shown in Figure~\ref{fg:overview} is presented in Figure~\ref{fg:temp}. The evolution from the room temperature to the sub-Verwey spectra occurs in several stages. In the first one, in the range between 250~K and 115~K, mostly the shape of the high-temperature peaks is affected. For example, a shoulder appears on the right hand side of the $T_{2g}(3)$ mode. Upon further cooling, new peaks appear at 160 and 480~cm$^{-1}$, the  $T_{2g}(3)$ mode at 310 cm$^{-1}$ breaks into several peaks and a shoulder appears at the left side of the $A_{1g}$ mode. 

There are no reported first-principle studies of the modes using the atomic positions of the trimeron model\cite{SennNature2012}. However, some of the changes can be understood qualitatively. For example, while there is only a single tetrahedral site in the high temperature cubic phase, in the monoclinic structure\cite{SennNature2012} there are eight different tetrahedral sites. A study\cite{GasparovPRB2000} of the expected modes has been done with a simpler orthorhombic unit cell with $Pmca$ symmetry, in which there is a doubling of the unit cell along one axis while the other axis are the diagonals of the cubic cell. In this case 78 Raman active modes are expected, instead of the 5 found for the cubic phase. So obviously many more peaks should, and in fact do appear. 

Some modes should persist with little changes across the Verwey transition. The cubic $A_{1g}$ mode is a prime example. In the low temperature phase, it corresponds to a $A_g$ mode, with which it shares the same Raman tensor. Other modes arise from the structural change from the cubic to the monoclinic unit cell. The $T_{2g}$ modes should split into $B_{1g}+B_{2g}+B_{3g}$ modes in the monoclinic phase. The modes arising from the high temperature cubic ones have been named "shoulder" modes\cite{YazdiNJP2013}. Such is the peak at 615~cm$^{-1}$. These modes are then appropriate to follow the structural changes in the unit cell of magnetite. And in addition, there are modes which are expected to arise due to the onset of the charge ordering below the Verwey transition, as has been observed for other oxides\cite{SchmidtPRL2003}. Those modes are unrelated to any of the cubic phase. Their origin is attributed to a substantial electrical polarization arising from off-centre atomic displacements in the trimeron charge-ordered phase. The modes at 160 and 480 cm$^{-1}$ belong to this group. They reflect, not the structural modifications but the electronic changes that correspond to the appearance of the charge-ordered state. 

Thus we separate the Raman peaks into three groups: those that do not show a large change across the transition, those that split into new peaks (which reflect the change in the unit cell), and the new peaks that arise from the appearance of the charge-ordered state. We discuss the evolution of peaks belonging to each group separately in order to take advantage of the power of Raman spectroscopy to track the different changes occurring in magnetite through the Verwey transition. In thin films, structural changes have been observed to precede the charge-order appearance when lowering the temperature\cite{YazdiNJP2013}.

\begin{figure}[htb]
	\centerline{\includegraphics[width=0.45\textwidth]{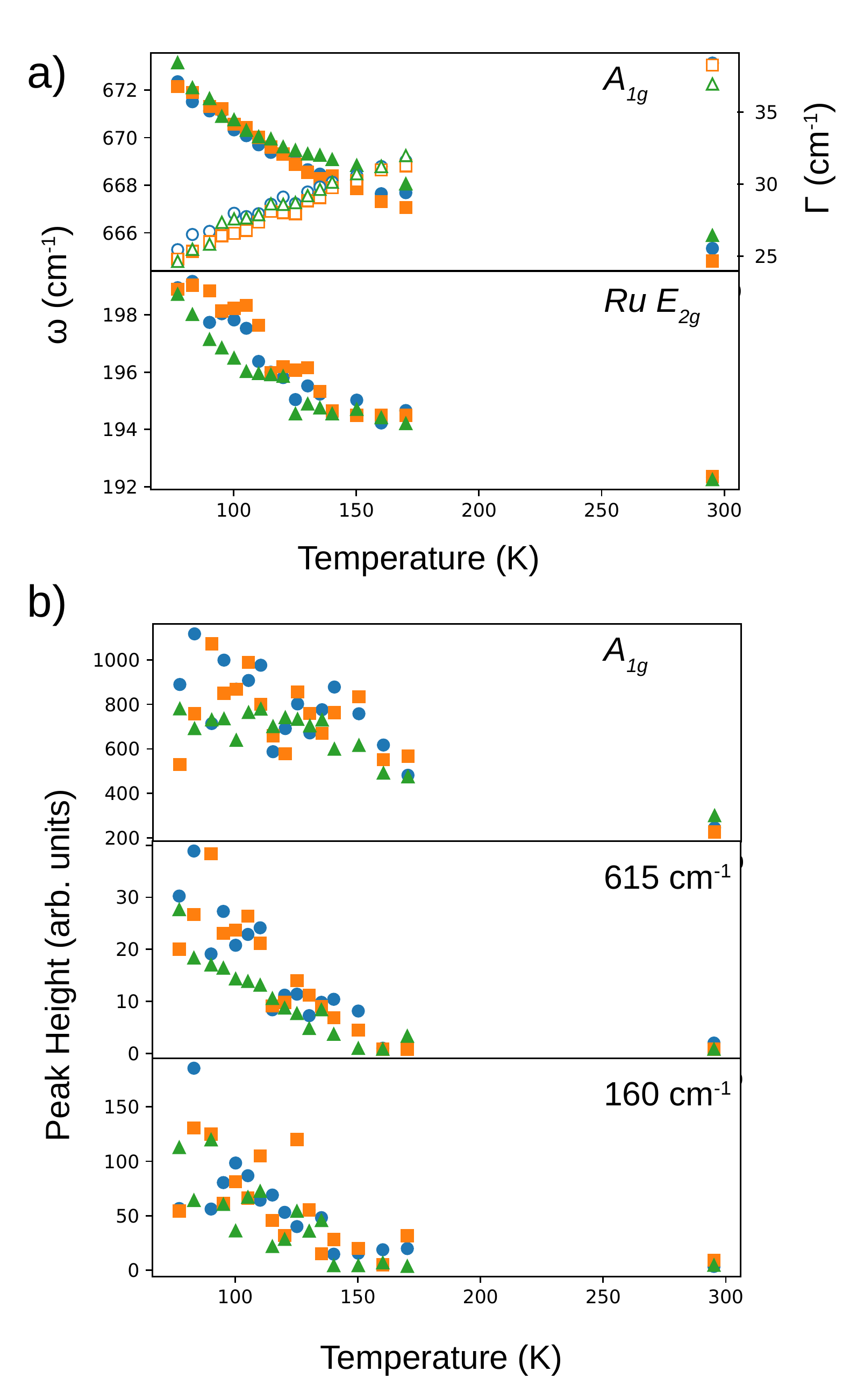}}
	\caption{Evolution with temperature of a) (top) $\omega$ and $\Gamma$ of the $A_{1g}$ mode and (bottom) $\omega$ of the Ru $E_{2g}$ mode. b) Peak height of the $A_{1g}$ Raman mode (top), the mode at 615~cm$^{-1}$ (middle), and the 160~cm$^{-1}$ one (bottom), respectively. The blue circles, orange squares and green triangles correspond to different islands which are  38 $\pm$ 5 nm, 37 $\pm$ 14 and 62 $\pm$ 20 nm thick, respectively. }
	\label{fg:evolution}
\end{figure}

The most intense mode is the cubic $A_{1g}$, which belongs to the first group. As Verble\cite{VerblePRB1974} already indicated, it does not drastically change across the Verwey transition, but smoothly changes its position and width (see Figure~\ref{fg:evolution}a). The evolution is similar to that reported for a high quality thin film\cite{YazdiNJP2013}. Other works report sharp changes in its width and position\cite{PhaseJAP2006,GasparovPRB2000}, which we do not observe. It is worth noting that the evolution of $A_{1g}$ is very similar to that of the Ru film mode.  

The $T_{2g}$ modes behave differently: their shape clearly changes in the range between 150~K and 115~K, reflecting the onset of a structural transition. This is nicely shown also by the shoulder mode at 615~cm$^{-1}$\cite{YazdiNJP2013}. As seen in Figure~\ref{fg:evolution}b, the increase in intensity at this energy starts already at 150~K.

The modes in the third group, such as the one at 160 cm$^{-1}$, which signal the onset of the charge-ordered state, do not appear until the temperature is lowered further (see Figure~\ref{fg:evolution}b, bottom). We interpret that the Verwey temperature of our crystals corresponds to the appearance of these latter peaks, following Yazdi et al.\cite{YazdiNJP2013}. As in their case, this implies that the structural changes precede the electronic ones upon lowering the temperature.

\begin{figure}[htb]
	\centerline{\includegraphics[width=0.45\textwidth]{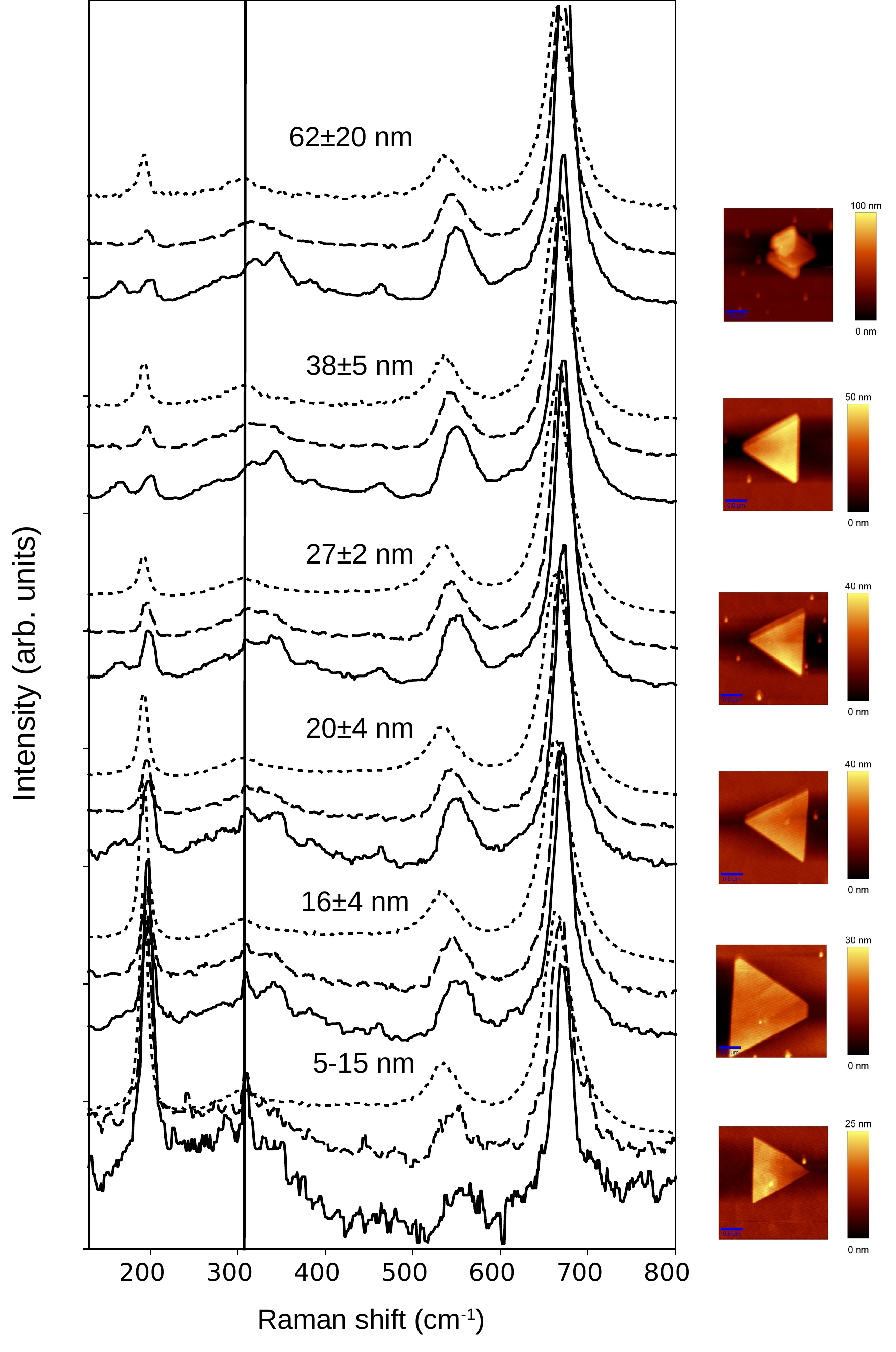}}
	\caption{Raman spectra of islands of different thickness, at room temperature (dotted line), at 115~K (dashed line) and at 77~K (continuous line). Atomic force microscopy images of each island are shown at the right hand side. A line marks the location of the sharp feature observed on the thinner islands at low temperature.}
	\label{fg:height_and_lt}
\end{figure}

Finally, we discuss the full low temperature spectrum as a function of island height, as shown in Figure~\ref{fg:height_and_lt}. The spectra of islands with thicknesses above 20~nm show the same features.
In all cases, there is a smooth evolution down to 115~K; below, we find the sudden appearance of the modes attributed to the sub-Verwey charge order. However, for crystals thinner than 20~nm the behavior changes, and for the thinnest crystal (between 5~nm and 15~nm, depending on the crystal side) the low-temperature spectrum, while clearly different from the room temperature one, differs also completely from the reference spectra of sub-Verwey magnetite. There is a very sharp new mode that only appears at low temperatures at 300~cm$^{-1}$ (marked with a blue line in the Figure), which is already observed for a thickness of 20~nm. We note that the location of that mode corresponds to the room temperature $T_{2g}(3)$ one. However, while the room temperature mode is wide for all islands (as discussed above), the low temperature feature is rather sharp. As that mode corresponds to an asymmetric breathing mode which in one direction corresponds to motion of the oxygen cations within the (111) film plane, while in the other to motion in a direction making an angle to that plane, it is tempting to relate this observation to some confinement effect of the $T_{2g}(3)$ mode. However, we note that we have not observed at room temperature any confinement effect whatsoever, something that might be attributed to the large width arising from the high electron-phonon interaction strength. Another surprising detail is that the position of this feature corresponds to the room temperature mode, but not to the expected position at low temperature, as observed in thicker films. 

In any case we thus have shown that the charge order of the crystals is similar to that of bulk magnetite only for thicknesses larger than 20~nm. For thinner crystals the Raman spectra changes drastically from the high-temperature cubic ones, indicating that there is a phase transition which changes the properties of the magnetite crystal, even if the transition to the bulk-like trimeron charge-ordered state does not take place. Regrettably, we lack at present a microscopic characterization of the crystallographic structure of such islands. Work to determine such structure is planned.

\section{Summary}

We have measured the Raman spectra of magnetite micro-crystals with thicknesses of tens of nanometers. They present room temperature Raman spectra typical of bulk magnetite. Our estimates of the electron-phonon interaction strength are similar to the values for the best single-crystals and thin films reported in the literature. At 77~K the spectra of crystals thicker than 20~nm correspond to those reported for the low-temperature phase of bulk magnetite. With decreasing temperature, down to 115~K, first the high-temperature phase peaks corresponding to the $T_{2g}$ modes change in shape. We do not observe any abrupt change in the $A_{1g}$ mode. Below 115~K, peaks corresponding to the charge order in the low-temperature phase of magnetite appear. However, for crystals thinner than about 20~nm, our results suggest that although a transition to a new phase takes place, its charge order and structure differ from the bulk case, highlighting the role of size effects in the Verwey transition.   

\section{Methods}

The Ru films were deposited on Al$_2$O$_3$(0001) single crystal substrates in a home-built magnetron sputtering system. A typical DC magnetron power used was 20~W after 10~min pre-sputtering of a 2" Ru target from Evochem GmbH. The substrates were kept at 900~K during film growth. A detailed characterization of the films by Rutherford Backscattering Spectroscopy and channeling has been recently described\cite{PrietoASS2022}. The films were then transferred to an ultrahigh vacuum chamber containing a low-energy electron microscope and annealed at temperatures of up to 1300~K. Both the Elmitec III LEEM at the IQFR and the Elmitec SPELEEM at the CIRCE beamline of the ALBA Synchrotron\cite{CIRCE} were used for the growth of the magnetite islands.  The growth was performed by introducing a pressure of 10$^{-6}$ mbar of molecular oxygen while the substrate was kept at a temperature of 800$^\circ$C and Fe was deposited from a home-made doser consisting of a Fe rod 5~mm in diameter in a water jacket heated by electron bombardment from a W filament with a typical power of 25~W. X-ray based characterization using photoemission microscopy was performed at the  ALBA Synchrotron CIRCE beamline.

The Raman spectra were acquired with a commercial Renishaw Witec Alpha 300RA confocal Raman spectrometer, using a 20$\times$ objective with a numerical aperture of 0.4. The light source was a 532~nm laser operated at 1~mW power, selected in order to avoid oxidation of the samples. The spectra presented are the average of 5 scans, each acquired with 30~s integration time. Measurements were performed at different temperatures between room temperature and liquid nitrogen temperature (77~K). The focus was adjusted at each temperature. We have fitted the Raman spectra by a sum of Lorentzians including a third degree polynomial for the background after applying a fifth order median filter. 

\begin{acknowledgments}

This work is supported by the Grants RTI2018-095303-B-C51, -A-C52 and -B-C53  funded by MCIN/AEI/10.13039/501100011033 and by “ERDF A way of making Europe”, and by the Grant  S2018-NMT-4321 funded by the Comunidad de Madrid and by “ERDF A way of making Europe”. S. Ruiz-Gomez thanks the Alexander von Humboldt Foundation for financial support.

\end{acknowledgments}

\bibliography{VerweyRaman}

\end{document}